%% file: main.tex
\documentclass[sigconf]{acmart}
\AtBeginDocument{%
  }

\copyrightyear{2026}
\acmYear{2026}
\setcopyright{cc}
\setcctype{by-nc-nd}
\acmConference[CHI EA '26]{Extended Abstracts of the 2026 CHI Conference on Human Factors in Computing Systems}{April 13--17, 2026}{Barcelona, Spain}
\acmBooktitle{Extended Abstracts of the 2026 CHI Conference on Human Factors in Computing Systems (CHI EA '26), April 13--17, 2026, Barcelona, Spain}
\acmDOI{10.1145/3772363.3798531}
\acmISBN{979-8-4007-2281-3/2026/04}

\usepackage{color}
\usepackage{soul}    
\usepackage{gensymb} 
\usepackage{subcaption}
\usepackage{xcolor}
\usepackage{graphicx}
\usepackage{enumitem} 
\usepackage{tabularray}
\usepackage{listings}
\usepackage{gensymb} 
\usepackage{xspace} 
\usepackage{pifont}
\usepackage{tabularx}
\usepackage{multirow}
\usepackage{array}
\usepackage{ragged2e}

\newcommand{\systemName}{\textsc{XAgen}\xspace}

\lstdefinestyle{prompt}{
  basicstyle=\ttfamily\small\color{black},
  showstringspaces=false,
  breaklines=true
}
\definecolor{mybrown}{HTML}{7B241C}
\definecolor{myblue}{HTML}
{0E6655}
\definecolor{mygray}{HTML}{595959}

\begin{document}

\title[\textsc{XAgen}: An Explainability Tool for Identifying and Correcting Failures in Multi-Agent Workflows]{\systemName: An Explainability Tool for Identifying and Correcting Failures in Multi-Agent Workflows
}



\author{Xinru Wang}
\orcid{0000-0002-0213-6425}
\affiliation{%
  \institution{Singapore-MIT Alliance for Research and Technology}
  \country{Singapore}}
\email{xinru.wang@smart.mit.edu}
\authornote{This work was done primarily while the author was an intern at Adobe Research and a PhD student at Purdue University.}

\author{Ming Yin}
\orcid{0000-0002-7364-139X}
\affiliation{%
  \institution{Purdue University}
  \city{West Lafayette}
  \country{IN, USA}}
\email{mingyin@purdue.edu}

\author{Eunyee Koh}
\orcid{0000-0003-2091-5972}
\affiliation{%
  \institution{Adobe Research}
  \city{San Jose}
  \country{CA, USA}}
\email{eunyee@adobe.com}

\author{Mustafa Doga Dogan}
\orcid{0000-0003-3983-1955}
\affiliation{%
  \institution{Adobe Research}
  \city{Basel}
  \country{Switzerland}}
\email{doga@adobe.com}

\renewcommand{\shortauthors}{Wang et al.}

\begin{abstract}
As multi-agent systems powered by Large Language Models (LLMs) are increasingly adopted in real-world workflows, users with diverse technical backgrounds are now building and refining their own agentic processes. However, these systems can fail in opaque ways, making it difficult for users to observe, understand, and correct errors. We conducted formative interviews with 12 practitioners to identify mismatches between existing debugging tools and users’ needs. Based on these insights, we designed \systemName, an explainability tool that supports users with varying AI expertise through three core capabilities: log visualization for glanceable workflow understanding, human-in-the-loop feedback to capture expert judgment, and automatic error detection via an LLM-as-a-judge. In a user study with 8 participants, \systemName helped users locate failures more easily, attribute to specific agents or steps, and iteratively improve configurations. Our findings surface human-centered design guidelines for explainable agentic AI development and highlight opportunities for more context-aware interactive debugging.
\end{abstract}


\begin{teaserfigure}
\centering
  \includegraphics[width=\textwidth]{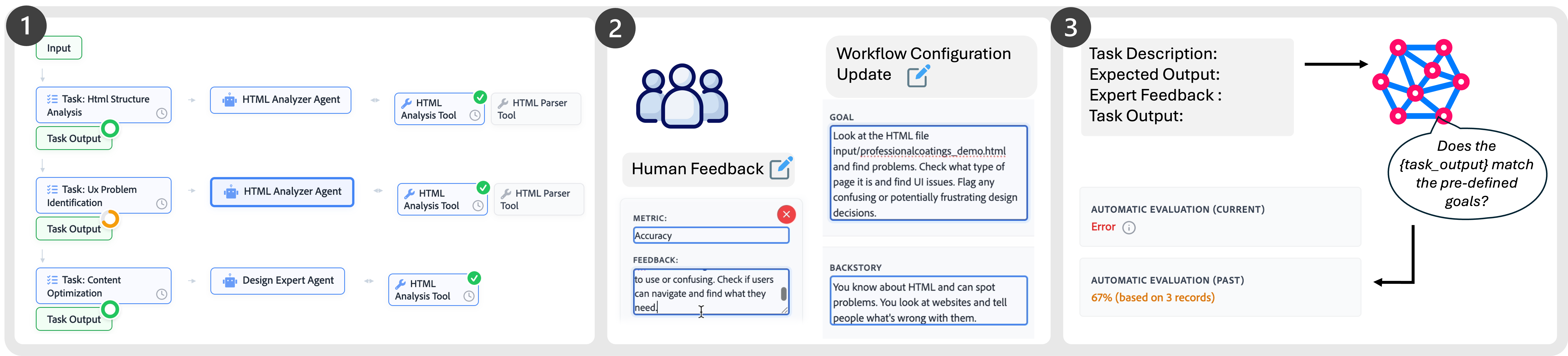}
  \caption{We introduce \systemName, an explainability tool that helps users with diverse AI expertise understand, detect, and correct failures in multi-agent workflows. \systemName\ integrates three core capabilities: (1) log visualization, which parses complex raw terminal logs into an interactive flowchart of agents, tasks, and tools, with each component in the flowchart activated step by step in accordance with the logs; 
  (2) human-in-the-loop feedback, which enables users to review outputs, provide feedback, and iteratively update agent and task configurations directly in the interface before re-running the workflow to test improvements;
  and (3) automatic error identification, which uses an LLM-as-a-judge to evaluate task outcomes against predefined goals and any collected user feedback, with the historical success rates shown as a ring indicator on the node, and detailed scores and rationales displayed in the details panel.
  }
  \label{fig:teaser}
\end{teaserfigure}


\maketitle

\section{Introduction}
\input{tex/01-introduction}

\section{Related Work}
\input{tex/02-relatedwork}

\input{tex/03-formativestudy}

\input{tex/04-system}

\input{tex/05-userstudy}

\input{tex/06-discussion}

\input{tex/07-conclusion}

\begin{acks}
We would like to thank our colleagues at Adobe Research and the \textit{Adobe LLM Optimizer} and \textit{AEM Sites Optimizer} teams for their thoughtful feedback and support throughout this project.
We are also grateful to all participants who contributed their time to our formative study and user study. Finally, we thank all anonymous reviewers for their insightful and constructive comments. 
This research has been partially supported by the National Research Foundation (NRF), Prime Minister’s Office, Singapore, under its Campus for Research Excellence and Technological Enterprise (CREATE) program, through the Mens, Manus and Machina (M3S) interdisciplinary research group (IRG) of the Singapore–MIT Alliance for Research and Technology (SMART) center.
\end{acks}

\balance
\bibliographystyle{ACM-Reference-Format}
\bibliography{main}

\appendix
\onecolumn
\input{tex/08-appendix}









\end{document}

%% file: tex/01-introduction.tex
Large Language Models (LLMs) have accelerated the democratization of multi-agent systems in a wide range of domains, from website optimization~\cite{he_webvoyager_2024, rieder_simab_2026}, co-creation~\cite{liu_how_2024, drago_improvmate_2025}, to content editing~\cite{yang_swe-agent_2024, tatsukawa_fontcraft_2025}. Unlike traditional monolithic AI models, multi-agent systems allow for the orchestration of specialized AI agents with distinct capabilities to solve complex tasks~\cite{guo_large_2024,fourney_magentic-one_2024}.
In practice, users with diverse AI expertise—from professional developers to designers and analysts—are now involved in configuring and maintaining these multi-agent workflows~\cite{ding_frontend_2025,ding_designgpt_2023}. For these users, a central challenge to effectively optimize multi-agent workflows is to \emph{identify and correct failure cases} over time: understanding when something has gone wrong and how to intervene. 

A common way to inspect a multi-agent workflow today is by reading and visualizing raw execution logs. These logs, which contain detailed traces of agent activities, communications, and tool usage, are typically viewed either in terminal windows or through developer-oriented observability dashboards such as AgentOps~\cite{agentopsai_agentops_2025}, LangTrace~\cite{langtrace_langtrace_2025}, and LangFuse~\cite{finto_technologies_langfuse_2025}. While powerful, these dashboards are primarily designed for AI developers with substantial technical expertise and focus on monitoring infrastructure. Users with less technical backgrounds, such as domain experts who increasingly collaborate in building multi-agent systems, may need more accessible support to more easily comprehend system behavior and contribute meaningful corrections.

We argue that making multi-agent workflows more usable requires \textit{importing and extending ideas from \textbf{Explainable AI}} \textbf{(XAI)}. XAI focuses on making model predictions more interpretable to human users for appropriate error detection and human intervention, e.g., by approximations~\cite{lundberg_unified_2017}, counterfactuals~\cite{wachter_counterfactual_2018}, or rationales~\cite{wei_chain--thought_2022}. However, unlike traditional single-model systems, failures in multi-agent systems are often distributed and non-obvious. For instance, a content-generation workflow may produce an incoherent final output not because a single agent malfunctioned, but because an upstream summarization agent omitted critical context, which then propagated through downstream rewriting and formatting agents. Rather than explaining a single model prediction, explaining multi-agent systems should involve tracing \textit{\textbf{dynamic system-level behaviors} between \textbf{multiple agents} and \textbf{tasks}}.

In this work, we explore how to design an explainability tool that helps users \textit{with varying AI expertise levels} identify and correct failure cases in multi-agent workflows. Drawing on formative interviews with 12 practitioners who actively work with multi-agent systems, spanning \textit{AI developers who \textbf{build} agentic systems} to \textit{domain experts who \textbf{utilize} agent workflows} in practice, we designed \systemName. The interface centers on three representative capabilities that jointly support error detection and correction (\autoref{fig:teaser}):

\begin{itemize}
\item \textbf{Log visualization for glanceable workflow understanding.} Instead of leaving users to parse raw logs, \systemName turns execution traces into an interactive flowchart. Each task, agent, and tool is represented as a node, giving users a holistic, glanceable view of how the workflow unfolded.
\item \textbf{Human-in-the-loop feedback to capture expert judgment.} \systemName allows users to annotate intermediate and final outputs with their own comments at any step.
\item \textbf{Automatic error identification with an LLM as a judge.} \systemName integrates an LLM to evaluate task outcomes against predefined goals and accumulated human judgments. These signals help users automatically detect potential failures or low-quality results.

\end{itemize}

We conducted a preliminary user evaluation with 8 participants who worked on two multi-agent system improvement tasks, one for webpage design and another for academic writing, using both \systemName\ and a baseline consisting of raw logs plus a commercial observability dashboard. Our results show that \systemName\ helped users more efficiently \textbf{locate failures}, \textbf{attribute them to specific agents or steps}, and \textbf{iteratively improve configurations}. Participants also surfaced open challenges, such as the need for \textit{richer intermediate visualizations}.

Taken together, our work makes the following two key contributions: (1) empirically grounded design guidelines for making multi-agent workflows \textbf{\textit{explainable}} for users with \textit{diverse} AI expertise, and (2) \systemName\ as a concrete system that supports error detection and correction in multi-agent workflows.

%% file: tex/02-relatedwork.tex
\textbf{Explainable AI for LLMs and Multi-Agent Systems.}
The black-box nature of emerging AI models can lead to misconceptions~\cite{wang_are_2021,wang_effects_2022}, bias~\cite{wang_effects_2023}, as well as reduced trust and task success~\cite{wang_human-llm_2024,wang_less_2025}. These issues have motivated both practitioners and researchers to pursue XAI approaches that help human users understand the internal reasoning of AI models. 
Traditional XAI techniques include feature attribution~\cite{ribeiro_why_2016,lundberg_unified_2017}, example-based explanations such as prototypes~\cite{kim_examples_2016} and counterfactuals~\cite{wachter_counterfactual_2018}, as well as inherently interpretable models such as rule-based systems and generalized additive models~\cite{jung_simple_2017,lakkaraju_interpretable_2016}. With LLMs, these techniques face new challenges due to massive training data, emergent capabilities, and non-deterministic outputs~\cite{dierk_evaluating_2025}. A common way to explain LLM responses is through “self-reasoning’’ such as chain-of-thought~\cite{wei_chain--thought_2022}, ReAct~\cite{yao_react_2023}, and multimodal reasoning~\cite{zhang_multimodal_2024}. 

While early work has pioneered XAI techniques for \textit{individual} LLMs, explainability remains an open challenge for multi-agent systems. Multi-agent systems are inherently more complex: multiple agents perform diverse roles, trigger tool use, exchange messages, and coordinate over time. These interactions generate overwhelming amounts of text, which makes it challenging for human users (e.g., a developer) to evaluate the factuality and quality of results~\cite{epperson_interactive_2025}.  

\textbf{Observability and Visual Analytics Tools for Multi-Agent Systems.}
The goals of explainability span multiple dimensions: beyond transparency, explanations should support error detection, enable human intervention and correction, and ultimately foster more effective human–AI collaboration~\cite{wang_are_2021,wang_effects_2022}. 

In software engineering, and increasingly in agentic AI systems, the term \textit{observability} refers to debugging tools that enable developers to monitor a system’s internal state during operation using external signals such as logs, traces, and metrics. In the context of AI agents, observability supports performance monitoring (e.g., latency, resource usage, response time) as well as error diagnosis and troubleshooting.
Most Agentic AI frameworks (e.g., CrewAI~\cite{crewai_crewaiinccrewai_2025}, AutoGen~\cite{microsoft_microsoftautogen_2025}, and LangGraph~\cite{langchain_langchain-ailanggraph_2025} provide detailed debugging logs of workflows and agent communications. Beyond these frameworks, commercial \textit{observability} platforms, such as AgentOps~\cite{agentopsai_agentops_2025}, Langfuse~\cite{finto_technologies_langfuse_2025} and Langtrace~\cite{langtrace_langtrace_2025}, provide visualization dashboards for execution traces and technical metrics such as time, tokens used, and error rates. 
Recent work on interactive debugging tools~\cite{epperson_interactive_2025,zhang_convomap_2025,zhang_chainbuddy_2025,holter_uxcascade_2026} further extends this by allowing users to edit agent behavior on the fly, and visualize multi-agent threads and state updates. For instance, Yan et al.~\cite{yan_answering_2025} provide a step-through visualization of the exploration path of code navigation agents, which complements existing output-centric dashboards. 

However, most existing observability tools are geared toward \textit{developers}. Other user groups---such as designers, product managers, and domain experts---who need to diagnose and correct issues in multi-agent systems may require different levels of transparency and support~\cite{suresh_beyond_2021,ehsan_who_2024}. We argue that extending XAI to multi-agent systems is crucial for aiding such users understand system behavior and correct issues. A user-centered design approach to explanation is therefore needed to align tools with the goals of users with diverse AI expertise levels.

%% file: tex/03-formativestudy.tex
\section{Formative Interviews with Practitioners}

To better understand how real-world practitioners improve multi-agent systems in their daily work, and the challenges they face with existing debugging methods, we conducted formative interviews with 12 practitioners recruited from our multinational organization. The sample included 7 developers, 2 research scientists, 1 engineering manager, and 2 domain experts with limited AI expertise. 
During the interview, they walked us through how they used agentic AI in a concrete example and reflected on challenges in understanding and debugging agents, and desired design features of tools that improve their debugging experience. All interviews were recorded, transcribed, and analyzed using thematic analysis. Below, we report representative themes raised by \textit{both technical and non-technical} participants. The full coding scheme is provided in \autoref{app:interview}.

\subsection{Challenges in Understanding AI Agents}
\textcolor{mybrown}{\textbf{C1. Steep learning curve and onboarding difficulties.}} 
A recurring challenge practitioners highlighted was the difficulty of using current interfaces and logs effectively. All participants described a steep learning curve and onboarding difficulties. For instance, P5 described \textit{“a very steep learning curve and a lot of frustration just to get it working in my local environ”}, while P11 estimated that \textit{“somebody will have to spend one week to install, to onboard”}.

\textcolor{mybrown}{\textbf{C2. Hallucination or incorrect model outputs.}}
Another recurring challenge arose from the behavior of LLMs themselves and their inherent limitations. Participants highlighted hallucinations and incorrect outputs as a common barrier. As P7 put it, \textit{“LLMs can hallucinate… they might produce an answer, but it may not be the best one, or even the right one.”} P11 observed that agents often continue executing even when outputs are clearly wrong: \textit{“even if something is wrong, the agents are very good at just going through—it will basically keep doing it, even though nothing went right.”}

\textcolor{mybrown}{\textbf{C3. Lack of reliable metrics and the dependence on domain expertise to interpret output quality.}}
Participants also described challenges in evaluating outputs and tracking the performance of agentic workflows, especially the lack of reliable metrics and the dependence on domain expertise to interpret output quality. P2 explained that \textit{“it’s very hard to achieve meaningful results using general-purpose LLMs. We should train our agents on domain-specific data… in our work we do have domain experts like SEO strategists, and they can manually evaluate results.”} 

\textcolor{mybrown}{\textbf{C4. Agent pipeline cannot be interrupted, need to track individual agent contributions.}}
Participants emphasized that challenges extend beyond individual agents to the end-to-end execution pipeline. Both developers and domain experts without a strong technical background noted that once a workflow begins, the agent pipeline cannot be interrupted, stressing the need to track individual agent contributions. As P1 explained, \textit{“With so many end-to-end tasks and agents, one flow takes too much time to finish before I can see the result and go back to improve—sometimes I wait ten minutes just to check a fix.” } When describing a three-agent loop, P7 asked, \textit{“Which agent should get what credit for this particular task? What is the best way to do credit attribution? We still don't know the answer.” } 

\subsection{Desired Features for Explainability Tools}

\textcolor{myblue}{\textbf{G1. Clean visualization of logs.}}
A recurring design goal was the need for a clean visualization of logs, including clear indications of the agent name, task, inputs/outputs, and communications. P5 emphasized that \textit{“the special agent that I created and the role and so on… having that sort of schema set up somewhere, like ‘oh, this agent is asking and this agent is answering,’ really helps.”} P2 similarly noted the value of seeing the overall flow: \textit{“If you have five agents, say like a manager coordinating a coding task in a software company, you can see where the process goes and how iterations happen… if you click on each of the nodes you can expand the input and output, and that would be really helpful.”} 

\textcolor{myblue}{\textbf{G2. Agent-level failure attribution.}}
Participants stressed the importance of being able to pause and isolate parts of a workflow. 
One highly requested feature was agent-level failure attribution, so that both novices and experts could better understand which agents were responsible for successes or failures. P7 noted that failures often trace back to a single agent, and being able to isolate its behavior would make troubleshooting more manageable. P3 added that: \textit{“Maybe if we have five agents interacting, only two actually made the change and the rest were just checking around… it’s interesting to measure the correlation between individual agent performance and group performance.”}

\textcolor{myblue}{\textbf{G3. Output check.}}
Participants emphasized the importance of systematic evaluation and tracking mechanisms to better understand agent and workflow performance. One desired capability was output checks—both immediate detection of invalid inputs/outputs and downstream/intermediate performance metrics. Such checks were seen as helpful for AI novices as well as more experienced users. P3 added that correlating activity metrics (latency, steps, tool usage) with real-world outcomes could provide deeper insights into agent efficiency.

%% file: tex/04-system.tex
\begin{figure*}[htbp]
  \centering
\includegraphics[width=\linewidth]{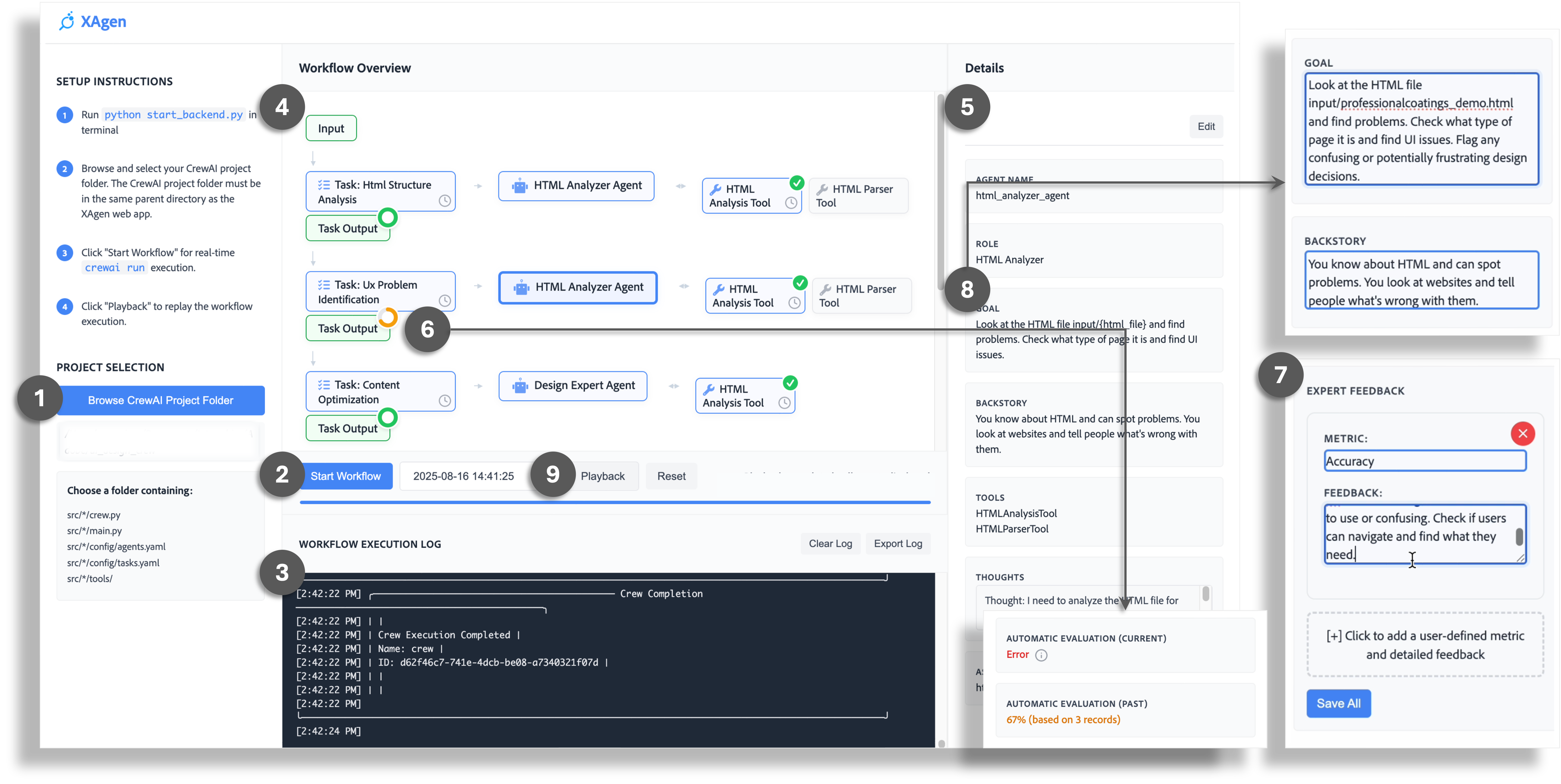}
  \caption{Walkthrough of the \systemName\ interface. Users first select a project folder (\textcolor{mygray}{\ding{202}}). We use the CrewAI framework in this prototype, as practitioners in our organization who participated in the formative interview primarily relied on CrewAI in their daily work. The upper section of the central panel displays the workflow overview, and clicking Start Workflow (\textcolor{mygray}{\ding{203}}) executes the multi-agent workflow. The lower section of the central panel then shows the raw terminal logs (\textcolor{mygray}{\ding{204}}). During execution, each component in the flowchart is activated step by step in accordance with the log (\textcolor{mygray}{\ding{205}}). The right panel displays detailed information, including prompt configurations, tool calls, and agent rationales (\textcolor{mygray}{\ding{206}}). Task outputs are automatically evaluated by the LLM-as-a-judge (\textcolor{mygray}{\ding{207}}); the average historical evaluation score is shown as a ring, while detailed scores and rationales are listed in the right panel. The details panel also provides fields for manual feedback (\textcolor{mygray}{\ding{208}}). Users can edit prompt configurations directly in the interface (\textcolor{mygray}{\ding{209}}) and re-run the workflow to test improvements. Finally, each session can be replayed, allowing users to review historical performance and feedback (\textcolor{mygray}{\ding{210}}).}
  \Description{Walkthrough of the XAgen interface. The figure shows three main panels. On the left, the setup panel allows the user to select a CrewAI project folder and start the workflow. In the center, the workflow overview displays a flowchart of tasks, agents, and tools, which are activated step by step as the workflow executes. Below the flowchart, a terminal window displays raw execution logs. On the right, the details panel shows information such as agent names, roles, prompts, rationales, and evaluation results. Task outputs are automatically evaluated by an LLM-as-a-Judge, with results shown using color-coded indicators and historical success rates. The panel also provides fields for manual expert feedback, where users can add custom metrics and comments. Users can edit configurations and re-run workflows, and past runs can be replayed to review historical performance and feedback.}
  \label{fig:screenshot}
\end{figure*}

\section{\systemName: Explainable Agentic AI Tool}

We implemented three core features in \systemName: (1) log visualization, (2) human-in-the-loop feedback, and (3) automatic error identification. Below, we describe how users interact with each feature. \autoref{fig:screenshot} presents a full screenshot and annotated walkthrough of the interface. The system implementation details are provided in~\autoref{app:architecture}.

\textbf{Log Visualization.}
The log visualization feature was designed to address usability challenges related to onboarding and interpreting raw logs (\textcolor{mybrown}{C1}) and was guided by practitioners’ desiderata for quickly gaining an overview of system behavior (\textcolor{myblue}{G1}).
As illustrated in~\autoref{fig:screenshot}, the visualization represents agents, tasks, and tool usage as clean, structured blocks. This design enables users to follow the workflow at a glance and serves as a foundation for locating errors at specific steps in the pipeline. During execution, components in the flowchart are activated sequentially as new log events arrive. The system also supports replaying past workflow states.

\textbf{Human-in-the-Loop Output Feedback.}
This feature aligns with practitioners’ desiderata to validate intermediate results and provide human feedback (\textcolor{myblue}{G3}). It addresses both the challenge of relying on domain expertise to interpret output quality (\textcolor{mybrown}{C3}) and the risks posed by hallucinated or incorrect model outputs (\textcolor{mybrown}{C2}).
When users identify suspicious components in the workflow, they can review and provide feedback through an interface panel linked to the corresponding workflow nodes, as shown in~\autoref{fig:screenshot}. Users may also directly modify configuration files by editing relevant fields within the interface. Importantly, this human feedback is incorporated into the LLM-as-a-Judge component to automatically surface potential issues.

\textbf{Automatic Error Identification.}
This feature helps users localize problems without re-running entire pipelines (\textcolor{myblue}{G2}, \textcolor{myblue}{G3}), addressing the difficulty of tracing errors in long, non-interruptible workflows (\textcolor{mybrown}{C4}). 
As illustrated in~\autoref{fig:screenshot}, once a task completes and produces an output, the system applies an LLM-as-a-Judge approach to compare the output against predefined goals and human feedback. The evaluation prompt template used in our study is provided in~\autoref{app:llm_judge_prompt}. Users can access the LLM judge’s rationale by clicking the information icon. To help users monitor performance over time, the system computes moving averages across workflow executions and displays visual indicators on the flowchart to reflect aggregated success rates.

%% file: tex/05-userstudy.tex
\section{User Evaluation}

\begin{figure}[htbp]
  \centering
  \begin{subfigure}[b]{0.45\textwidth}
  \includegraphics[width=\linewidth]{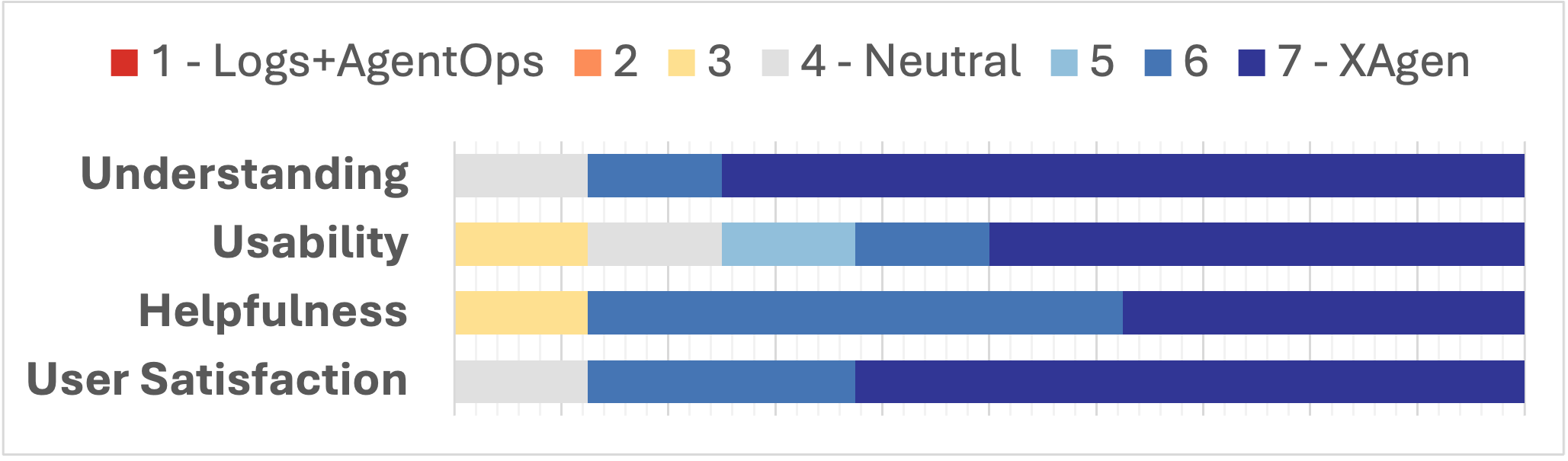}
        \caption{}
        \label{fig:result_all}
    \end{subfigure}
    \hspace{10pt}
    \begin{subfigure}[b]{0.45\textwidth}
\includegraphics[width=\linewidth]{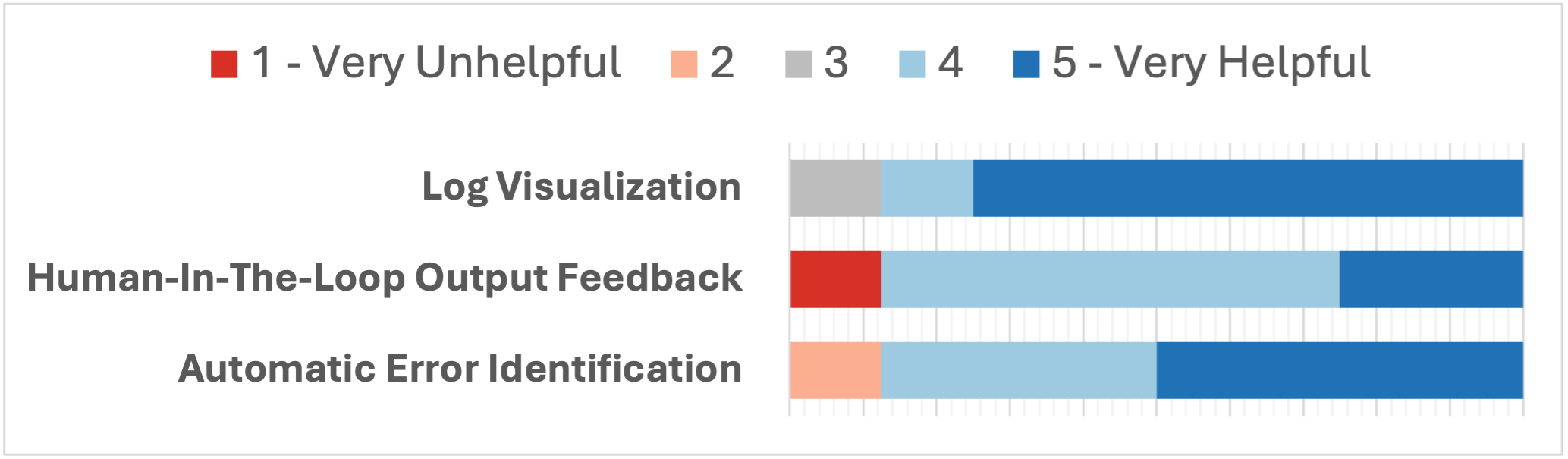}
        \caption{}
        \label{fig:result_individual}
    \end{subfigure}
    \caption{User study results comparing \systemName against a baseline (a) and evaluating helpfulness of \systemName's three core features (b).}
    \Description{User study results shown as two grouped bar charts.
(a) Ratings comparing XAgen with Logs+AgentOps across four dimensions: understanding, usability, helpfulness, and user satisfaction. Bars are color-coded from red (Logs+AgentOps, score 1) to dark blue (XAgen, score 7). The results show a clear preference for XAgen, with most ratings clustered at 6–7 across all four dimensions.
(b) Ratings of the helpfulness of XAgen's three core features: log visualization, human-in-the-loop output feedback, and automatic error identification. Bars are color-coded from red (1 = very unhelpful) to dark blue (5 = very helpful). Log visualization and automatic error identification were rated consistently as helpful (mostly 4–5), while human-in-the-loop output feedback received more mixed ratings, with some users finding it less helpful.}
  \label{fig:result}
\end{figure}

We invited 8 participants from our organization (mean age = 28, SD = 1.6) to preliminarily evaluate our system design. Participants reported diverse levels of prior experience in developing AI systems (5-point Likert scale, 1 = none, 5 = extensive): 4 rated themselves as ``extensive'', 1 as ``none'', 2 as ``a little'', and 1 as ``some''. We designed two tasks to simulate realistic multi-agent workflows: In the \textit{UI design task}, participants worked with a multi-agent workflow to improve a poorly designed webpage, while in the \textit{academic writing task}, to generate a related work section and a summary table given a research topic.
The initial workflow was intentionally configured to produce poorly organized or even erroneous outputs (e.g., missing CSS in the rendered webpage, or only listing authors, years, and links in the related work table). Participants were asked to revise the workflow configuration files based on the failure cases they identified. Their goal was to guide the agents toward generating outputs that better satisfied the task requirements. 

We compared \systemName\ with a baseline condition that combined raw system logs and AgentOps, a state-of-the-art observability tool. 
In the baseline condition, participants were introduced to interpreting terminal logs and the web-based AgentOps interface. While the interface of AgentOps visualizes logs and tracks technical metrics such as runtime and token usage, it is primarily designed for developers and focuses on infrastructure-level monitoring.
The study followed a within-subjects design with counterbalanced task and system order.

At the end of the study, participants rated their experience with both systems across several dimensions, including understanding~\cite{wang_are_2021,zhang_convomap_2025}, usability~\cite{zhang_chainbuddy_2025}, helpfulness~\cite{epperson_interactive_2025}, user satisfaction~\cite{wang_watch_2023,zhang_chainbuddy_2025}, as well as the usefulness of \systemName's three individual features~\cite{epperson_interactive_2025,zhang_chainbuddy_2025}. In addition to the surveys, participants provided open-ended feedback in a short follow-up interview.

\textbf{Results.}
Participants rated \systemName more positively than debugging with terminal logs and AgentOps across all four subjective metrics (\autoref{fig:result_all}), and reported that \systemName's three core features were useful overall (\autoref{fig:result_individual}).

Among the three features, log visualization was the most favored. Participants reported that flowchart-style view helped them more easily understand the overall workflow during debugging. Several participants noted that workflow outputs were still presented as dense text, and suggested richer visualization of intermediate and final outputs.
The human-in-the-loop feedback feature received comparatively lower engagement and the lowest usefulness ratings in our study. This limitation was partly due to the study setting, as participants had only 20 minutes per session, which was not sufficient to fully iterate on the workflow.
The LLM judge feature was valued for its helpfulness in identifying which agent was responsible for an error and quickly directed users to potential problem areas, as noted by P2, P5, and P6. Beyond error localization, the LLM judge also provided rationales for its evaluations. Participants sometimes directly incorporated text from these rationales into prompt edits to improve workflow performance. P4 explained, \textit{“I think the judge's feedback is often quite reasonable—sometimes even more thorough than my own.”}

%% file: tex/06-discussion.tex
\section{Discussion}


\textbf{System Improvements.}
Our system can benefit from design extensions informed by user study feedback. A recurring need was to better visualize text-heavy intermediate and final outputs from agentic workflows. One direction is to incorporate visual indicators directly into the interface~\cite{wang_ui_2026}. For instance, bounding boxes on a webpage could highlight problematic areas and allow errors to be located more quickly~\cite{hoque_hallmark_2024}. Alternatively, descriptive tours of key snippets could provide a coherent narrative that traces an AI agent's exploration path~\cite{yan_answering_2025}.
Second, \systemName's human feedback module shows potential for supporting collaborative work. If deployed on a server-based shared platform, domain experts could conveniently review and intervene in workflow outputs. This collaborative potential warrants further investigation in multi-user contexts.
Finally, although the LLM judge provided rationales for its evaluations, the system currently lacks features such as auto-fill suggestions for prompt revisions, which could further support more efficient iteration.

While the current design showcases linear workflows, future iterations should support non-linear structures, such as loops, hierarchies, and parallel execution, to handle more complex real-world agentic collaborations.

\textbf{Future Evaluation Directions.}
Our preliminary user evaluation provides initial guidance for iterative system design but has several limitations that we plan to strengthen. We will add a technical evaluation to validate how well individual features perform and to provide clearer performance characterization. We will use an ablated version of \systemName as a baseline by systematically removing each feature to better isolate its contribution. Finally, we will incorporate objective measures (e.g., time to identify errors, correction accuracy) to make the evaluation more rigorous. 

%% file: tex/07-conclusion.tex
\section{Conclusion}
In this work, we extend the concept of XAI to multi-agent systems to assist users in identifying and correcting failures. Grounded in formative interviews with real-world practitioners with diverse AI expertise, we design \systemName, an explainability tool that integrates log visualization, human-in-the-loop feedback, and automatic error identification. A user study with 8 participants shows how these features support failure detection and correction while revealing open challenges. We envision this work as a step toward the human-centered design of explainability interfaces that enable even novice users to participate in refining multi-agent workflows as AI increasingly collaborates with humans.

%% file: tex/08-appendix.tex
\clearpage
\section{Complete Qualitative Themes from the Formative Interviews}
\label{app:interview}

\begin{table*}[htbp]
\centering
\small
\caption{Challenges in Understanding and Debugging AI Agents}
\label{tab:themes_quotes}
\setlength{\tabcolsep}{6pt}
\renewcommand{\arraystretch}{1.15}

\begin{tabularx}{\textwidth}{
  >{\RaggedRight\arraybackslash}p{0.12\textwidth}
  >{\RaggedRight\arraybackslash}p{0.25\textwidth}
  >{\RaggedRight\arraybackslash}X
}
\hline
\textbf{Themes} & \textbf{Sub-themes} & \textbf{Examples} \\
\hline

\multirow{3}{=}{\textbf{Interface and Usability Challenges}} &
Steep learning curve and onboarding difficulties &
\textit{``A very steep learning curve and a lot of frustration just to get it working in my local environ''}~(P5). \\
\cline{2-3}
& Insufficient logging information &
\textit{``We have to manually add print statements in the code to see what’s happening. If some logs are missing, we still have to go back and add them ourselves when debugging.''}~(P9). \\
\cline{2-3}
& Difficulty managing large numbers of agents &
\textit{``Right now, I run maybe four servers and can check different terminals to see where the error is. But if I had 50 agents, I cannot do this \ldots what is the right way to look at these errors?''}~(P7). \\
\hline

\multirow{3}{=}{\textbf{Model Behavior and LLM-Specific Limitations}} &
Non-deterministic execution flow &
\textit{``The most troublesome part is that the agent’s outputs are not deterministic---sometimes you get a good format, sometimes nothing useful at all, and it’s impossible to reproduce. That’s the most frustrating part.''}~(P8). \\
\cline{2-3}
& Hallucination or incorrect model outputs &
\textit{``LLMs can hallucinate\ldots they might produce an answer, but it may not be the best one, or even the right one.''}~(P7). \\
\cline{2-3}
& Limited scalability of LLM solutions across diverse use cases &
\textit{``There is a big tendency to tailor a solution for one particular problem. But when we try to scale---not just in terms of load but in terms of variety of use cases---it starts to break. Fixing one issue often introduces unpredictable problems in new use cases.''}~(P3). \\
\hline

\multirow{2}{=}{\textbf{Output Evaluation and Performance Tracking}} &
Difficulty tracking performance across the workflow &
\textit{``With time it’s hard to get all the data and keep them reliable\ldots the website changes, everything around changes, so it’s hard to crystallize it.''}~(P3). \\
\cline{2-3}
& Unreliable metrics; requires domain expertise to interpret output quality &
\textit{``It’s very hard to achieve meaningful results using general-purpose LLMs. We should train our agents on domain-specific data\ldots in our work we do have domain experts like SEO strategists, and they can manually evaluate results. But we don’t have an automated system for that.''}~(P2). \\
\hline

\multirow{2}{=}{\textbf{Code and Execution Pipeline Complexity}} &
Agent pipeline cannot be interrupted; need to track individual agent contributions &
\textit{``With so many end-to-end tasks and agents, one flow takes too much time to finish before I can see the result and go back to improve---sometimes I wait ten minutes just to check a fix.''}~(P1). \\
\cline{2-3}
& Issues stemming from the broader code pipeline (beyond agents) &
\textit{``We take the agent’s output, do some small post-processing in our code, and then pass it to another agent. The complicated and frustrating part is figuring out where the problem really is---whether the agent is wrong or our code logic is wrong. You end up having to repeatedly go back and check everything.''}~(P8). \\
\hline

\multirow{3}{=}{\textbf{Organizational and Operational Constraints}} &
Limited access to customer data &
\textit{``Finding relevant data is also a challenge because customer data we don’t have---the ones we can have are from third-party tools, which require approval and payment to use.''}~(P1). \\
\cline{2-3}
& Hosting concerns (e.g., data privacy and security) &
--- \\
\cline{2-3}
& Enterprise restrictions in adopting commercial tools (e.g., legal approval due to user data privacy) &
--- \\
\hline
\end{tabularx}
\end{table*}

\begin{table*}[htbp]
\centering
\small
\caption{Desired Features for Explainability Tools}
\label{tab:design_implications_quotes}
\setlength{\tabcolsep}{6pt}
\renewcommand{\arraystretch}{1.15}

\begin{tabularx}{\textwidth}{
  >{\RaggedRight\arraybackslash}p{0.15\textwidth}
  >{\RaggedRight\arraybackslash}p{0.15\textwidth}
  >{\RaggedRight\arraybackslash}X
}
\hline
\textbf{Themes} & \textbf{Sub-themes} & \textbf{Examples} \\
\hline

\multirow{1}{=}{\textbf{Interface improvements for easier onboarding and visibility}} &
Clean visualization of logs: agent name, task, inputs/outputs, communications &
\textit{``The special agent that I created and the role and so on\ldots having that sort of schema set up somewhere, like `oh, this agent is asking and this agent is answering,' really helps.''}~(P5). \\
\hline

\multirow{2}{=}{\textbf{Automated code optimization and log analysis}} &
Architecture-level optimization recommendations (e.g., time, agent usage) &
\textit{``Saying you spent too much money or time on this task, and suggesting how to optimize the prompt or merge multiple tool calls into one agent.''}~(P1). \\
\cline{2-3}
& Querying logs, log summarization &
\textit{``I can just say, `analyze this log, I suspect the issue is in this part of the code,' and the system could quickly scan the logs, narrow down the scope, and even suggest fixes.''}~(P8). \\
\hline

\multirow{3}{=}{\textbf{Breakpoint-based debugging and agent isolation}} &
Run from breakpoint &
\textit{``It would be really cool to have breakpoints. I could stop at a point and see the input and output so far in the pipeline. That way I’d know everything is working up to here, and then if I add a new agent I can see whether things break after that.''}~(P6). \\
\cline{2-3}
& Agent-level failure attribution &
\textit{``Maybe if we have five agents interacting, only two actually made the change and the rest were just checking around\ldots it’s interesting to measure the correlation between individual agent performance and group performance.''}~(P3). \\
\cline{2-3}
& Standardized protocols helpful for distributed unit testing &
\textit{``You could isolate a component, trigger individual agents from your own interface, and test them independently—almost like distributed unit testing.''}~(P11). \\
\hline

\multirow{3}{=}{\textbf{Evaluation and performance tracking}} &
Agent + tool performance + resource + time tracking; third-party tool errors &
\textit{``I need to know where time is being spent—what is taking away my time.''}~(P7). \\
\cline{2-3}
& Batch behavior tracking / profiling &
\textit{``We need to somehow track—was it better or not, what worked and what didn’t, and learn from that history.''}~(P2). \\
\cline{2-3}
& Output checking &
\textit{``If the agent did something weird, or if the tools failed when calling an API, we don’t really know whether everything worked correctly.''}~(P4). \\
\hline
\end{tabularx}
\end{table*}

\clearpage
\section{Implementation Details of the System Architecture}
\label{app:architecture}
\begin{figure*}[htbp]
  \centering
  \includegraphics[width=0.75\linewidth]{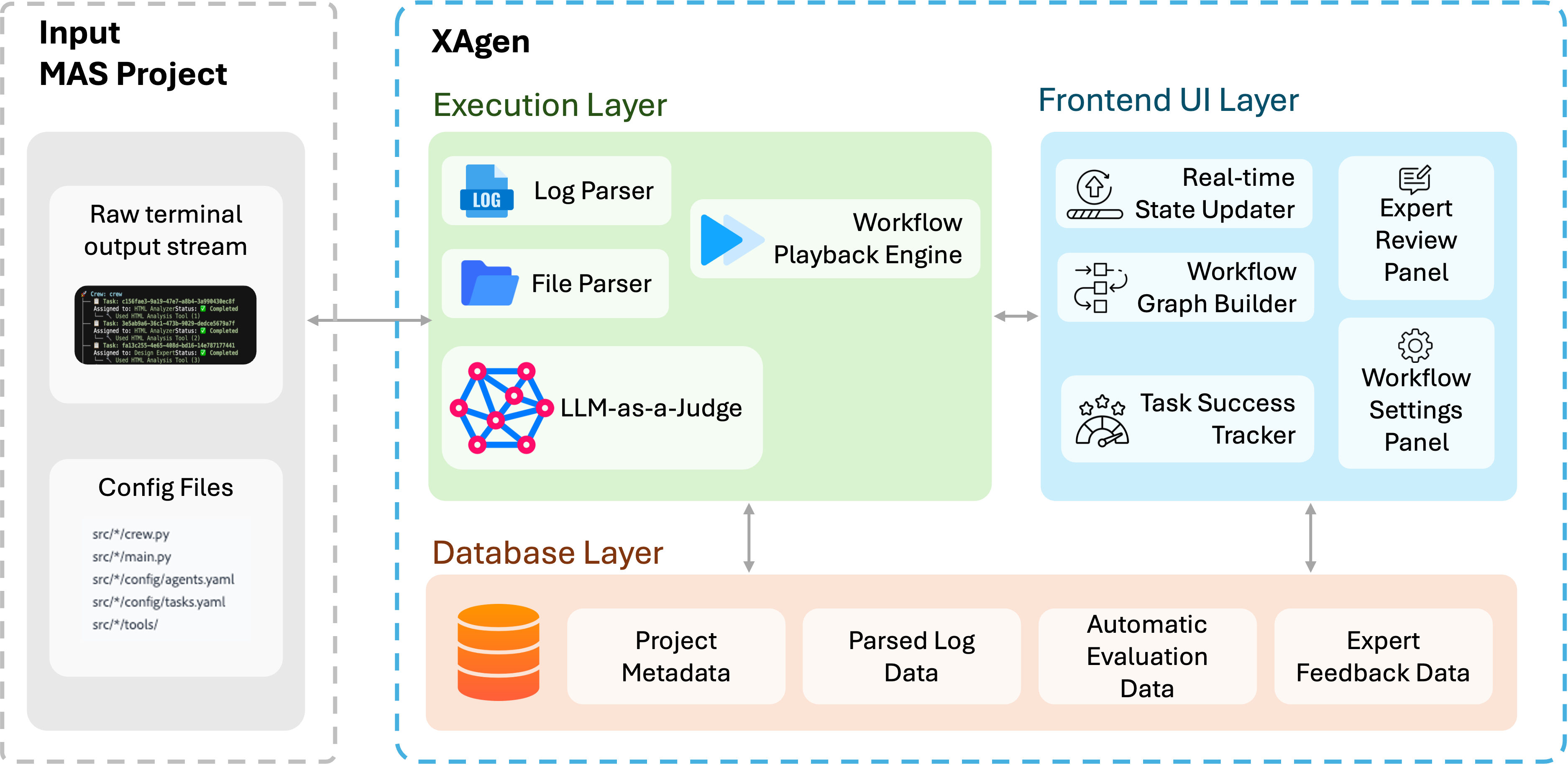}

  \caption{Architecture of \systemName.}
  \Description{System architecture of XAgen. The figure shows three layers: the execution layer (log parser, file parser, workflow playback engine, LLM-as-a-Judge), the frontend UI layer (state updater, workflow graph builder, task tracker, expert review panel, settings panel), and the database layer (project metadata, parsed logs, automatic evaluation data, expert feedback data). Arrows illustrate how execution logs and configuration files are processed, evaluated, and displayed to users. }
  \label{fig:architecture}
\end{figure*}

\autoref{fig:architecture} shows the technical implementation of \systemName. The system operates in parallel with any multi-agent project and consists of three layers. The Database Layer stores essential data required for restoration of past runs. The Execution Layer provides backend functionality such as parsing logs and configuration files, generating workflow visualizations, and using the LLM-as-a-Judge for automatic evaluation. The Frontend UI Layer presents this information to users. Edits made in the UI are transferred back to the execution layer for processing. 

The system reads CrewAI configuration files to construct dynamic workflow flowcharts. The system implements a multi-threaded log parsing architecture that processes CrewAI execution streams in real time. The core parser uses regex pattern matching to extract structured data from unstructured terminal output. 
Structured data and raw log entries are stored in JSON objects, then pushed to the frontend via WebSocket connections for real-time UI updates. Flowchart UI elements are dynamically activated as new log events arrive via WebSocket.

The system implements a database to store raw logs, extracted structures, and execution sessions. The playback engine reconstructs historical workflow states by replaying stored events. 

The interface provides an interactive UI for domain experts to review outputs and provide feedback. Feedback is linked to specific workflow components and stored in the database.
The system enables users to modify YAML configuration files directly through the interface by editing corresponding fields. Version control is handled via the database, which maintains timestamped backups of configuration files. 

The system includes an asynchronous evaluation pipeline that operates independently of the main workflow execution. When a task is completed and an output is generated, this triggers the evaluation pipeline. The system uses an LLM-as-a-Judge approach to compare task outputs against predefined goals and human feedback. We set the temperature to 0 to ensure more deterministic outputs. A scoring algorithm with uncertainty handling was applied: “yes” were scored as 1.0 (task meets requirements), “no” as 0.0 (task fails requirements), and “unsure” as 0.5 (ambiguous evaluation). In addition to these scores, the LLM judge also generated a rationale for each evaluation to support interpretability. Users can click on the information icon to access the detailed rationale provided by the LLM judge. To help users monitor performance over time, the system computes moving averages across workflow executions for historical tracking. Visual indicators are displayed on the flowchart interface to reflect the aggregated success rates.

\newpage
\section{LLM-as-a-Judge Prompt}
\label{app:llm_judge_prompt}
\begin{lstlisting}[style=prompt]
You are an impartial evaluator. Your task is to assess whether the FINAL ANSWER generated by an AI agent adequately satisfies the given TASK DESCRIPTION and EXPECTED OUTPUT, taking into account EXPERT FEEDBACK from previous attempts.
Please return your judgment in the following JSON format:
{{
  "label": "yes" | "no" | "unsure",
  "rationale": "<brief explanation of your reasoning>"
}}
Inputs:
TASK DESCRIPTION: {task_description} 
EXPECTED OUTPUT: {expected_output}
EXPERT FEEDBACK: {expert_feedback}
FINAL ANSWER: {final_answer}
Important: Respond with only the JSON object. Do not include any additional text or commentary.
\end{lstlisting}